\documentclass[]{emulateapj}

\newcommand {\apgt} {\ {\raise-.5ex\hbox{$\buildrel>\over\sim$}}\ }
\newcommand {\aplt} {\ {\raise-.5ex\hbox{$\buildrel<\over\sim$}}\ }

\shorttitle{LMC Cepheids from Spitzer Data}
\shortauthors{Ngeow \& Kanbur}

\begin{document}

\title{The Period-Luminosity Relation for the Large Magellanic Cloud Cepheids Derived from {\it Spitzer} Archival Data}

\author{C. Ngeow}
\affil{Department of Astronomy, University of Illinois, Urbana, IL 61801}
\and 
\author{S. M. Kanbur}
\affil{State University of New York at Oswego, Oswego, NY 13126}

\begin{abstract}

Using {\it Spitzer} archival data from the SAGE (Surveying the Agents of a Galaxy's Evolution) program, we derive the Cepheid period-luminosity (P-L) relation at $3.6$, $4.5$, $5.8$ and $8.0$ microns for Large Magellanic Cloud (LMC) Cepheids. These P-L relations can be used, for example, in future extragalactic distance scale studies carried out with the {\it James Webb Space Telescope}. We also derive Cepheid period-color (P-C) relations in these bands and find that the slopes of the P-C relations are relatively flat. We test the nonlinearity of these P-L relations with the $F$ statistical test, and find that the $3.6\mu \mathrm{m}$, $4.5\mu \mathrm{m}$ and $5.8\mu \mathrm{m}$ P-L relations are consistent with linearity. However the $8.0\mu \mathrm{m}$ P-L relation presents possible but inconclusive evidence of nonlinearity.

\end{abstract}


\section{Introduction}

The Cepheid period-luminosity (P-L) relation is an important component in extragalactic distance scale and cosmological studies. The most widely used P-L relations in the literature are obtained from Large Magellanic Cloud (LMC) Cepheids in optical $BVI$ \citep[e.g., see][]{mad91,tan99,uda99a,san04,kan06} and near infra-red (NIR) $JHK$ \citep[e.g., see][]{mad91,gie98,gro00,nik04,per04,nge05} bands. In addition, there are also some LMC P-L relations obtained for MACHO $V_{MACHO}R_{MACHO}$ \citep{nik04,nge05} and EROS $V_{EROS}R_{EROS}$ \citep{bau99} bands. Recently, \citet{nge07} applied a semiempirical approach to derive the LMC P-L relation in Sloan $ugriz$ bands: still within optical and/or NIR regimes. Therefore, Cepheid P-L relations are well-developed for wavelengths ranging from optical to NIR. 

In contrast, there are currently no P-L relations available for wavelengths longer than the $K$ band in the literature. The main motivation for having a P-L relation at these longer wavelengths is in order to apply it in future extragalactic distance scale studies. The {\it Near Infrared Camera (NIRCam)} and the {\it Mid-Infrared Instrument (MIRI)}, that are scheduled to be installed on the {\it James Webb Space Telescope (JWST)}, will operate in the NIR and mid-infrared: a wavelength range of 0.6-5 microns and 5-27 microns, respectively\footnote{See the links given in {\tt http://www.stsci.edu/jwst/instruments/}}. It is possible that extragalactic Cepheids will be discovered and/or (re-)observed (for those galaxies that were observed by the $HST\ H_0$ Key Project) by the {\it JWST}. Hence, P-L relations at longer wavelengths are needed in order to use {\it JWST} to derive a Cepheid distance to these galaxies. In this Paper, we derive the LMC P-L relations at $3.6$, $4.5$, $5.8$ and $8.0$ microns from the {\it Spitzer} archival data. As far as we are aware, this is the first time such a relation has been derived. 

In addition to the P-L relations, we can also derive period-color (P-C) relations and construct color-color plots and the color-magnitude diagrams (CMD) for LMC Cepheids in these {\it Spitzer} bands. In Section 2 we discuss our data selection. In Section 3 we present our analysis and results for the P-L relations. Section 4 we show the P-C relations, the CMD and the color-color plot for the Cepheids in our sample. Our conclusion is given in Section 5. Extinction is ignored in this Paper because it is expected to be negligible in the {\it Spitzer's IRAC} bands (hereafter {\it IRAC} band).

\section{Data Selection}

The SAGE \citep[Surveying the Agents of a Galaxy's Evolution,][]{mei06}\footnote{See {\tt http://sage.stsci.edu/index.php}} is a program to survey the LMC using the {\it Spitzer} satellite. As a result it has detected about 4 million sources in the LMC with {\it Spitzer's IRAC} instrument (with an angular resolution of $\sim2$ arcsecond). The data are publicly available via the IRSA's Gator Catalog Query\footnote{See {\tt http://irsa.ipac.caltech.edu/applications/Gator/}}. To find LMC Cepheids in the SAGE database, we first obtained the right ascension (RA) and declination (DEC) of the LMC Cepheids from the OGLE \citep[Optical Gravitational Lensing Experiment,][]{uda99b} database. We only obtained the Cepheids that were classified as {\it fundamental} mode Cepheids (labelled as FU) from the OGLE database, where the classification was mainly based on the $W_I$ P-L relation and the Fourier decomposition technique \citep{uda99b}. The initial list of 771 Cepheids was cross-correlated with a list of "good" Cepheids given in \citet{kan06}. The details regarding selection criteria used to remove ``bad'' Cepheids, including possible overtone Cepheids, can be found in \citet{kan06} and will not be repeated here. This left 627 OGLE LMC Cepheids. To increase the sample size and to extend the period coverage to longer period (OGLE Cepheids truncated at $P\sim30$ days due to CCD saturation), we added non-OGLE Cepheids from the \citet{seb02} catalog. As in previous studies \citep[for example, see][and the reference therein]{uda99a,kan06,nge05,san04}, we applied a period cut of $P>2.5$ days to our sample to avoid inclusion of overtone Cepheids. The number of Cepheids left in our sample is 737.

\begin{deluxetable}{lcc}
\tabletypesize{\scriptsize}
\tablecaption{Search Results from the SAGE Database.\label{tab0}}
\tablewidth{0pt}
\tablehead{
\colhead{} &
\colhead{SAGE Catalog} &
\colhead{SAGE Archive} }
\startdata
$N_{\mathrm{total}}$    & 886 & 912 \\
$N_{\mathrm{no match}}$ & 13  & 4   \\
$N_{\mathrm{final}}$    & 724 & 733 \\
$<d>$                    & $ 0.777\pm0.015$ & $0.773\pm0.014$ \\ 
                         & st. deviation$=0.393$ & st. deviation$=0.374$ \\
$<\Delta RA>$            & $ 0.575\pm0.020$ & $0.582\pm0.019$ \\
                         & st. deviation$=0.531$ & st. deviation$=0.518$ \\
$<\Delta DEC>$           & $-0.088\pm0.014$ & $-0.092\pm0.013$ \\
                         & st. deviation$=0.372$ & st. deviation$=0.349$ \\
\enddata
\tablecomments{Units for $<d>$, $<\Delta RA>$ and $<\Delta DEC>$ are in arcsecond.}
\end{deluxetable}

\begin{figure}
\plotone{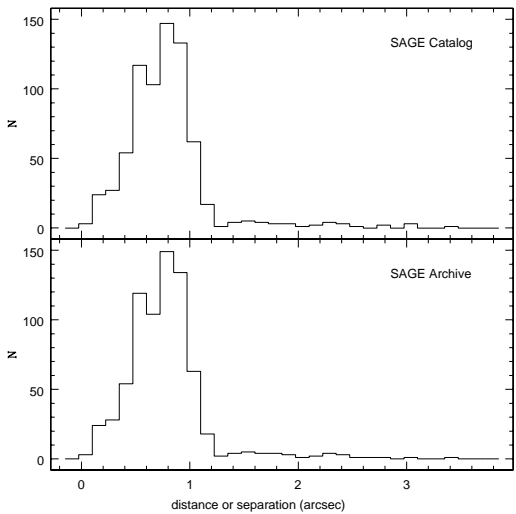}
\caption{The distribution of the ``distance'', or the separation, between the Cepheid locations and the locations of the matched sources. \label{fig_dist}}
\end{figure}

The IRSA's Gator Catalog Query provides two SAGE datasets: the SAGE Winter '07 IRAC Catalog (hereafter SAGE Catalog) and the SAGE Winter '07 IRAC Archive (hereafter SAGE Archive). The difference between the SAGE Catalog and the SAGE Archive can be found in \citet{mei06} and in the SAGE document\footnote{See {\tt http://irsa.ipac.caltech.edu/applications/Gator/\\ GatorAid/SAGE/SAGE\_SSCdatadocument\_delivered.pdf}\label{ftnote}}, and will not be repeated here. The search on the SAGE Catalog and Archive were performed with a search radius of $3.5$ arcseconds. The results are summarized in Table \ref{tab0}. In this table, $N_{\mathrm{total}}$ is the total number of matched sources. However there are several Cepheids which do not have any matched sources within the search radius: the number of these Cepheids is given as $N_{\mathrm{no match}}$. For those Cepheids with multiple matched sources, the sources with a minimum ``distance'' or separation from the input RA and DEC of the LMC Cepheids were selected to remove the duplicated SAGE sources. The number of matched sources left in the sample is summarized as $N_{\mathrm{final}}$ in Table \ref{tab0}. The mean ``distance'', $<d>$, between the input Cepheid locations and the locations of the matched sources, and the mean difference in RA and DEC, $\Delta RA$ and $\Delta DEC$ respectively, for the remaining Cepheids are presented in Table \ref{tab0} as well. Figure \ref{fig_dist} shows the distribution of the ``distance'' or the separation between the remaining Cepheids and the matched sources from both of the SAGE Catalog and the SAGE Archive.

Running the same query on SAGE Winter '07 MIPS $24\mu \mathrm{m}$ Catalog returned $12$ and $25$ matched sources with $5$ and $8$ arcsecond search radius, respectively. We therefore do not consider the detections from the MIPS catalog in this Paper.

\section{The Period-Luminosity Relation}

\begin{figure*}
\plottwo{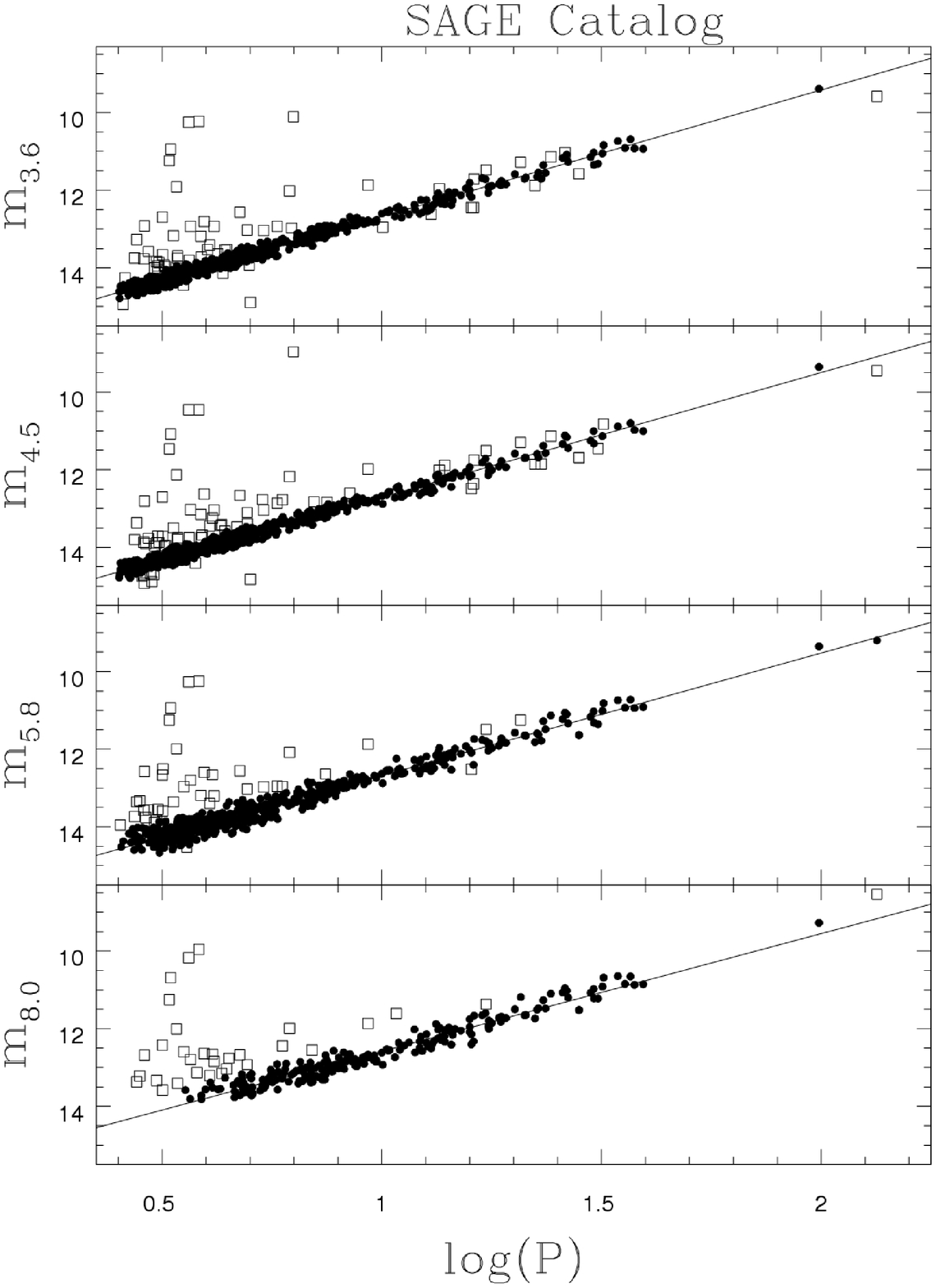}{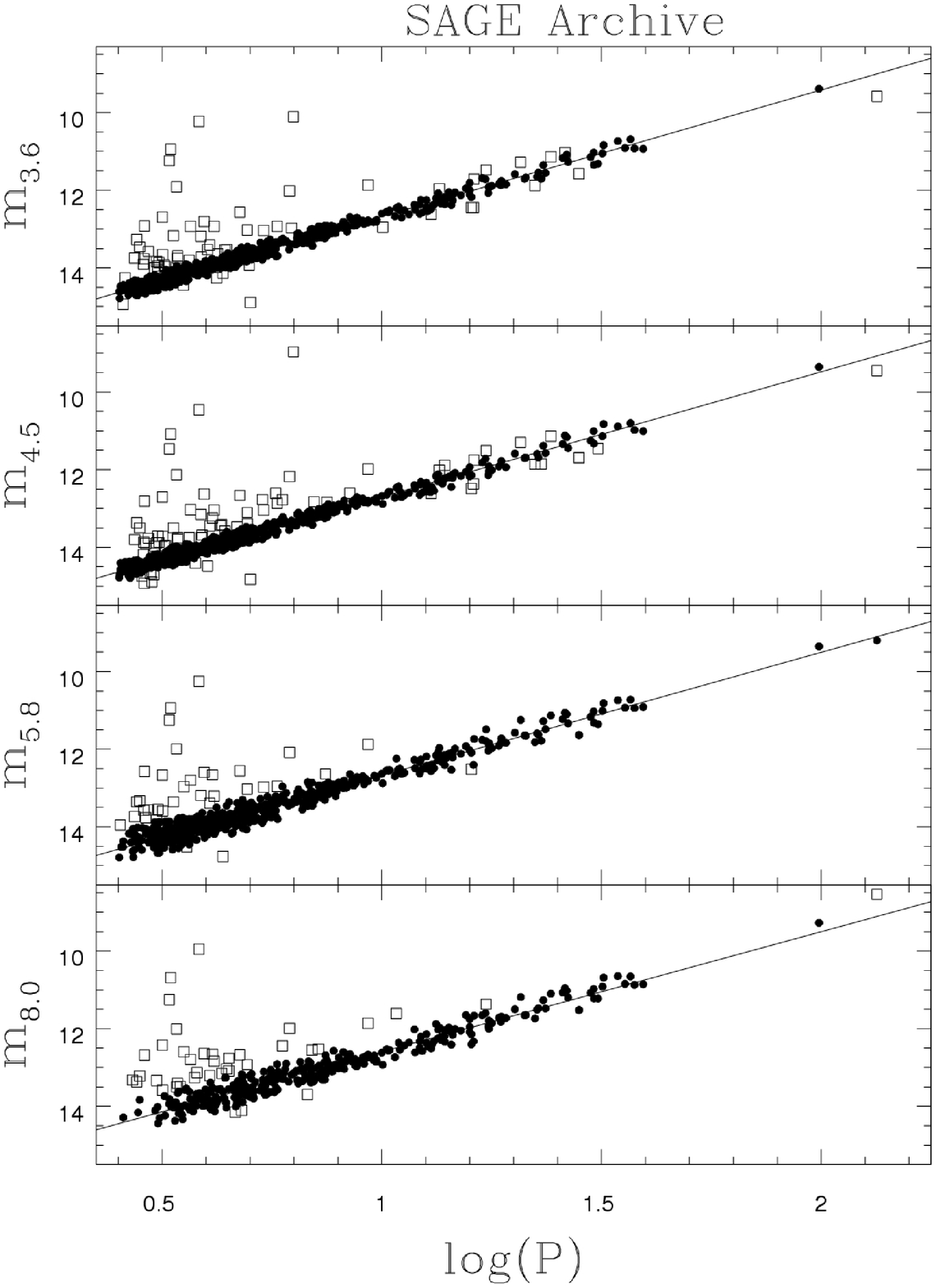}
\caption{The P-L relations from the matched sources in the SAGE Catalog (left panel) and the SAGE Archive (right panel), after an iterative process to remove the outliers. The open squares and solid circles are for the rejected outliers and the remaining data points, respectively. The lines are the fitted P-L relations to the data as given in Table \ref{tab1}. \label{fig_pl}}
\end{figure*}

Initial plots of the P-L relations for all matched sources display a tight P-L relation with some obvious outliers. These outliers are probably due to the mis-match of the Cepheids with the input SAGE database, blending of other sources along the line-of-sight, or other physical reasons. These reasons are difficult to track down because we only have the data from the publicly available database and not the source images. Therefore we apply an iterative outlier removal algorithm \citep[the sigma-clipping algorithm, as in][]{uda99a} to remove the outliers. For each iteration, P-L relations are fitted to the data and outliers located more than $2.5\sigma$ away from the fitted regression lines are removed, where $\sigma$ is the dispersion of the regression lines. This process is repeated several times until the solutions from the regression become stable. The rejected outliers are represented as open squares in Figure \ref{fig_pl}. 

\begin{deluxetable}{lcccc}
\tabletypesize{\scriptsize}
\tablecaption{P-L Relations in {\it IRAC} Band.\label{tab1}}
\tablewidth{0pt}
\tablehead{
\colhead{Band} &
\colhead{Slope} &
\colhead{Zero-Point} &
\colhead{$\sigma$} &
\colhead{$N$}}
\startdata
\cutinhead{SAGE Catalog} \\
$3.6\mu \mathrm{m}$  & $-3.265\pm0.017$ & $15.947\pm0.012$ & 0.104 & 613 \\
$4.5\mu \mathrm{m}$  & $-3.211\pm0.017$ & $15.922\pm0.012$ & 0.104 & 618 \\
$5.8\mu \mathrm{m}$  & $-3.158\pm0.027$ & $15.840\pm0.022$ & 0.169 & 534 \\
$8.0\mu \mathrm{m}$  & $-3.031\pm0.048$ & $15.609\pm0.049$ & 0.183 & 215 \\
\cutinhead{SAGE Archive} \\
$3.6\mu \mathrm{m}$  & $-3.263\pm0.016$ & $15.945\pm0.012$ & 0.104 & 628 \\
$4.5\mu \mathrm{m}$  & $-3.221\pm0.017$ & $15.927\pm0.012$ & 0.103 & 635 \\
$5.8\mu \mathrm{m}$  & $-3.173\pm0.028$ & $15.850\pm0.022$ & 0.175 & 561 \\
$8.0\mu \mathrm{m}$  & $-3.091\pm0.039$ & $15.684\pm0.036$ & 0.193 & 319 
\enddata
\tablecomments{$\sigma$ is the dispersion of the P-L relation.}
\end{deluxetable}

\begin{figure*}
\plottwo{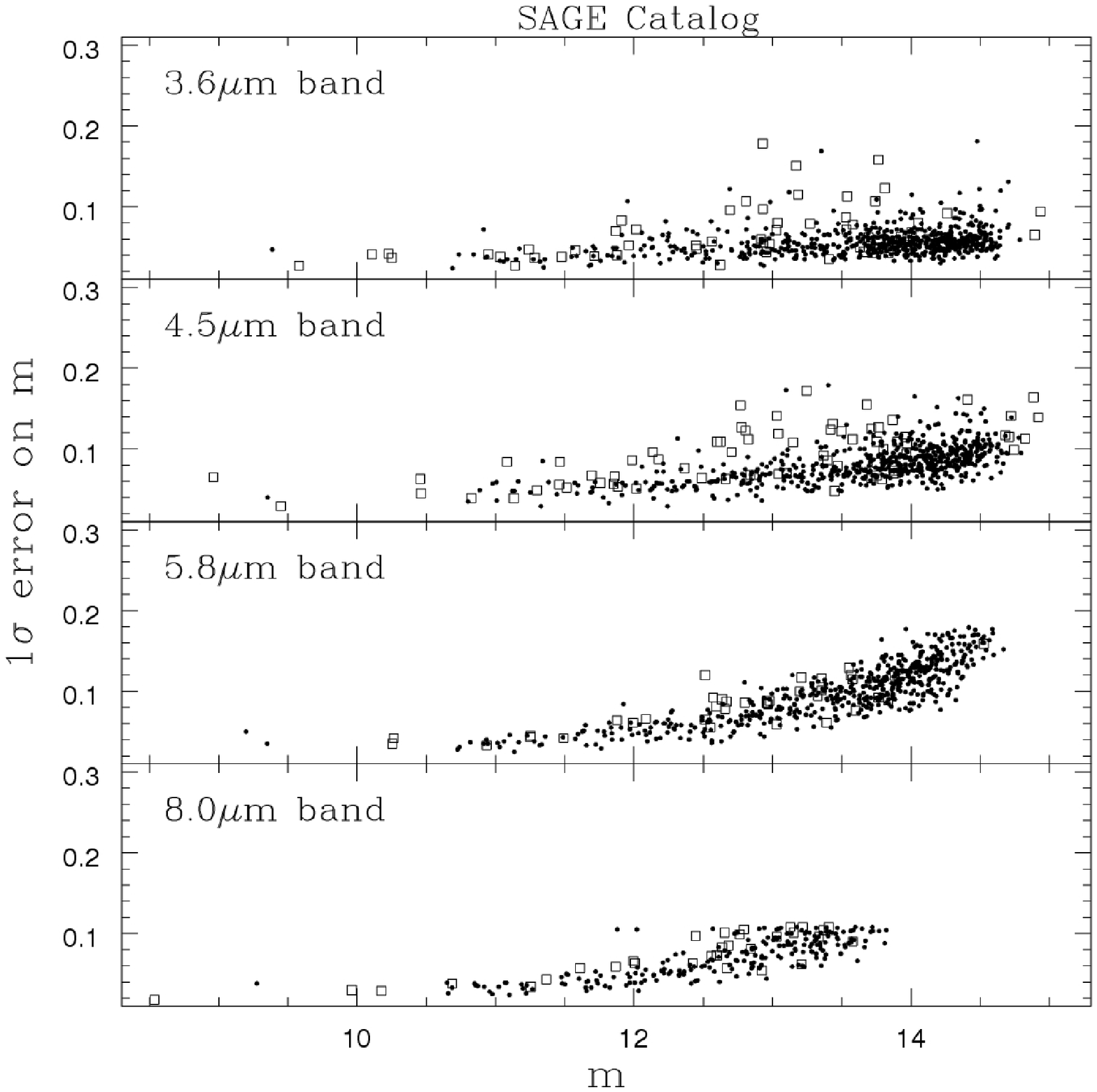}{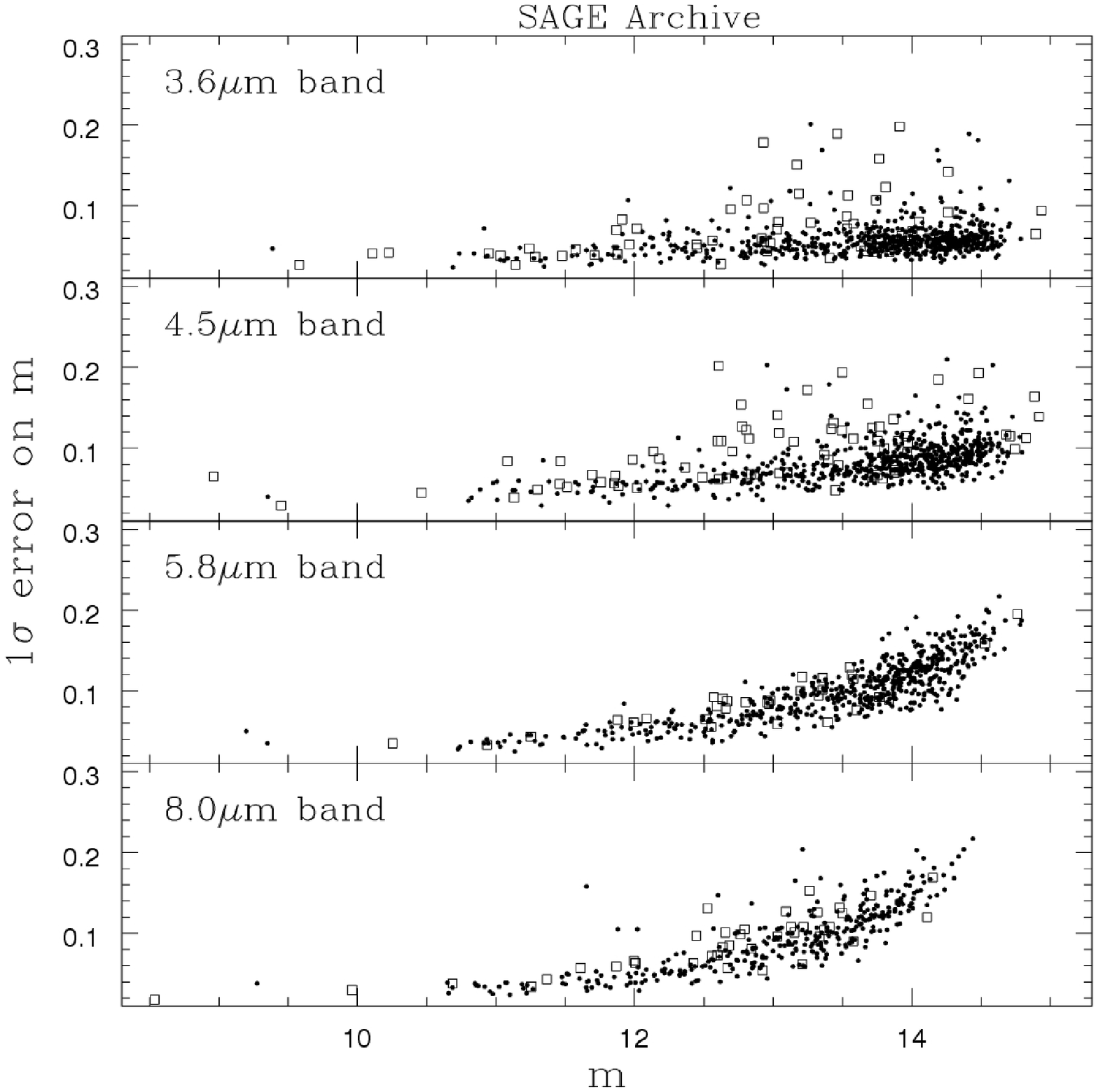}
\caption{Plots of the $1\sigma$ errors on the magnitudes as reported in the SAGE database versus the magnitudes. The open squares and solid circles are for the rejected outliers and the remaining data points, respectively.  \label{fig_mmerr}}
\end{figure*}

Figure \ref{fig_pl} displays the P-L relations in the {\it IRAC} bands and Table \ref{tab1} presents the results from fitted regression lines after outliers have been removed. When fitting the P-L relations, we do not stipulate constraints that the number of Cepheids should be the same in all four bands and/or a given Cepheid is detected in all four bands because the number of Cepheids in the $8.0\mu \mathrm{m}$ band is much smaller than in other bands. Table \ref{tab1} finds that the {\it IRAC} band  P-L relations obtained using data from the SAGE Catalog and the SAGE Archive are in very good agreement. A small discrepancy is seen for the $8.0\mu \mathrm{m}$ P-L relation, perhaps due both to the different numbers of Cepheids in the two catalogs (there are $\sim48$\% more Cepheids from the SAGE Archive than in the SAGE Catalog), and also the lack of Cepheids fainter than $m=14$mag. in the SAGE Catalog (see lower-left panel of Figure \ref{fig_pl}). In Figure \ref{fig_mmerr}, we give the reported $1\sigma$ errors on the magnitudes as a function of magnitude from the SAGE database. This figure clearly shows the truncation of $m=14$mag. at $8.0\mu \mathrm{m}$ for our Cepheid sample in the SAGE Catalog. Furthermore, there is a clear cut-off of the $1\sigma$ errors for the $8.0\mu \mathrm{m}$ band data\footnote{This cut-off is still visible if we include a large number of non-Cepheid data from the SAGE Catalog.} from the SAGE Catalog, which is absent in other panels in Figure \ref{fig_mmerr}. The lack of $m_{8.0}>14$mag. and the cut-off of the $1\sigma$ errors at $8.0\mu \mathrm{m}$ band could be due to the more stringent criteria for the SAGE Catalog than the SAGE Archive \citep[see][and the SAGE document given in footnote \ref{ftnote}]{mei06}. For these reasons, we only consider the data from SAGE Archive in the rest of this Paper.

\begin{figure}
\plotone{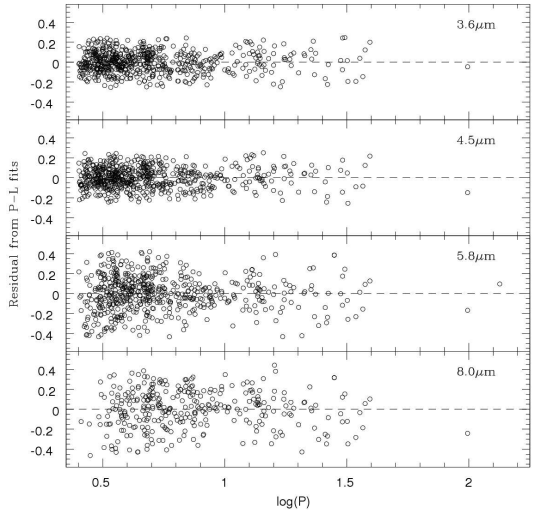}
\caption{The residuals from the P-L relations given in Table \ref{tab1} (SAGE Archive) as a function of period.\label{fig_plres}}
\end{figure}

Figure \ref{fig_plres} presents the residual plots from the fitted P-L relation in each bands. The residuals for $3.6\mu \mathrm{m}$, $4.5\mu \mathrm{m}$ and $5.8\mu \mathrm{m}$ band are more or less evenly distributed around the regression lines. However the residuals for $8.0\mu \mathrm{m}$ band exhibit a deficit on one side of the regression at $\log (P)\sim 1.0$. This could suggest that the $8.0\mu \mathrm{m}$ band P-L relation may not be linear, and this will be investigated further in Section \ref{sec33}. Interestingly, the smallest dispersion of the P-L relations seems to occur at $\log (P)\sim 1.0$ in all four bands.

\subsection{Sensitivity of the P-L Relations with Various Period Cuts}

\begin{figure*}
\plottwo{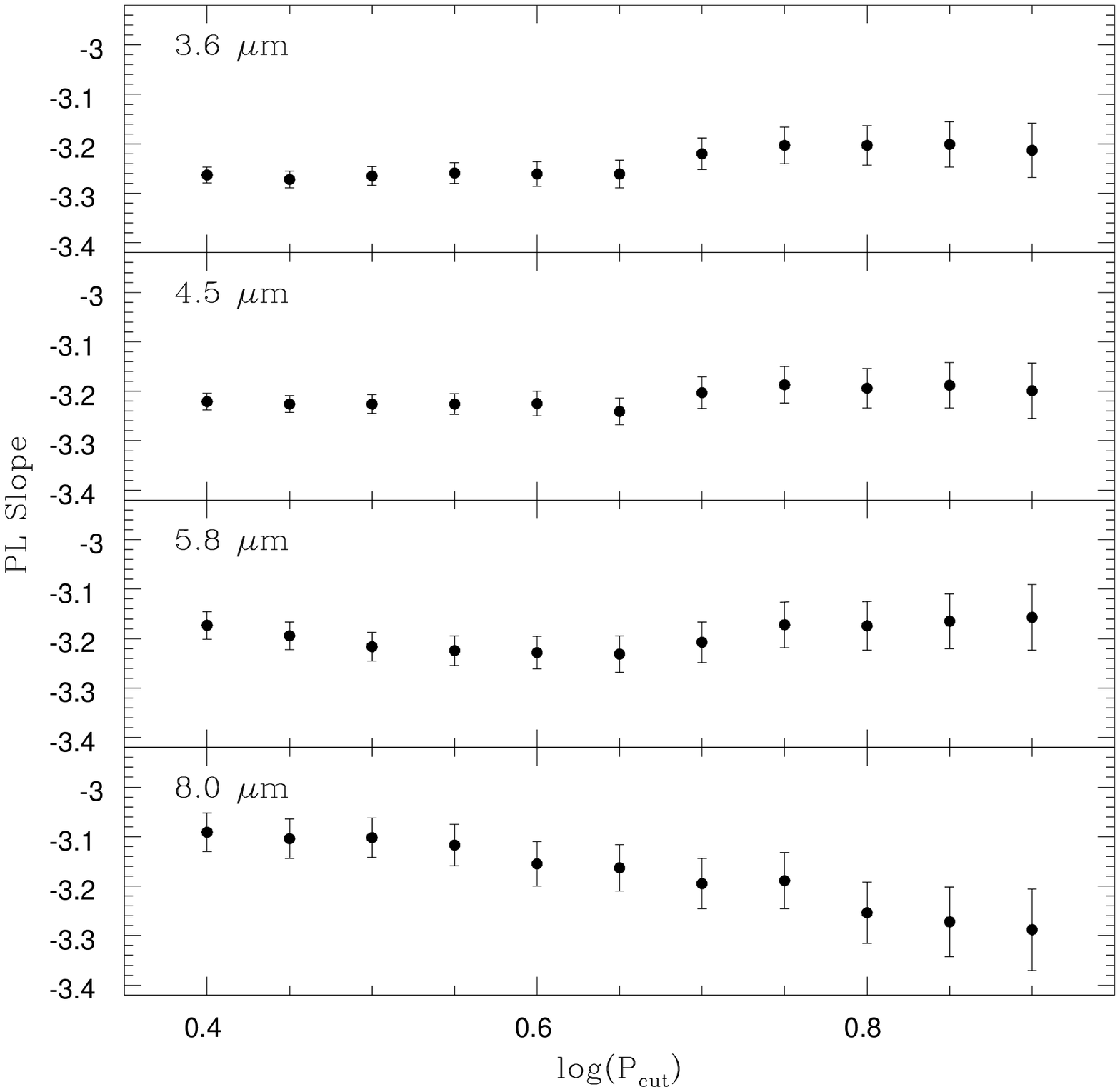}{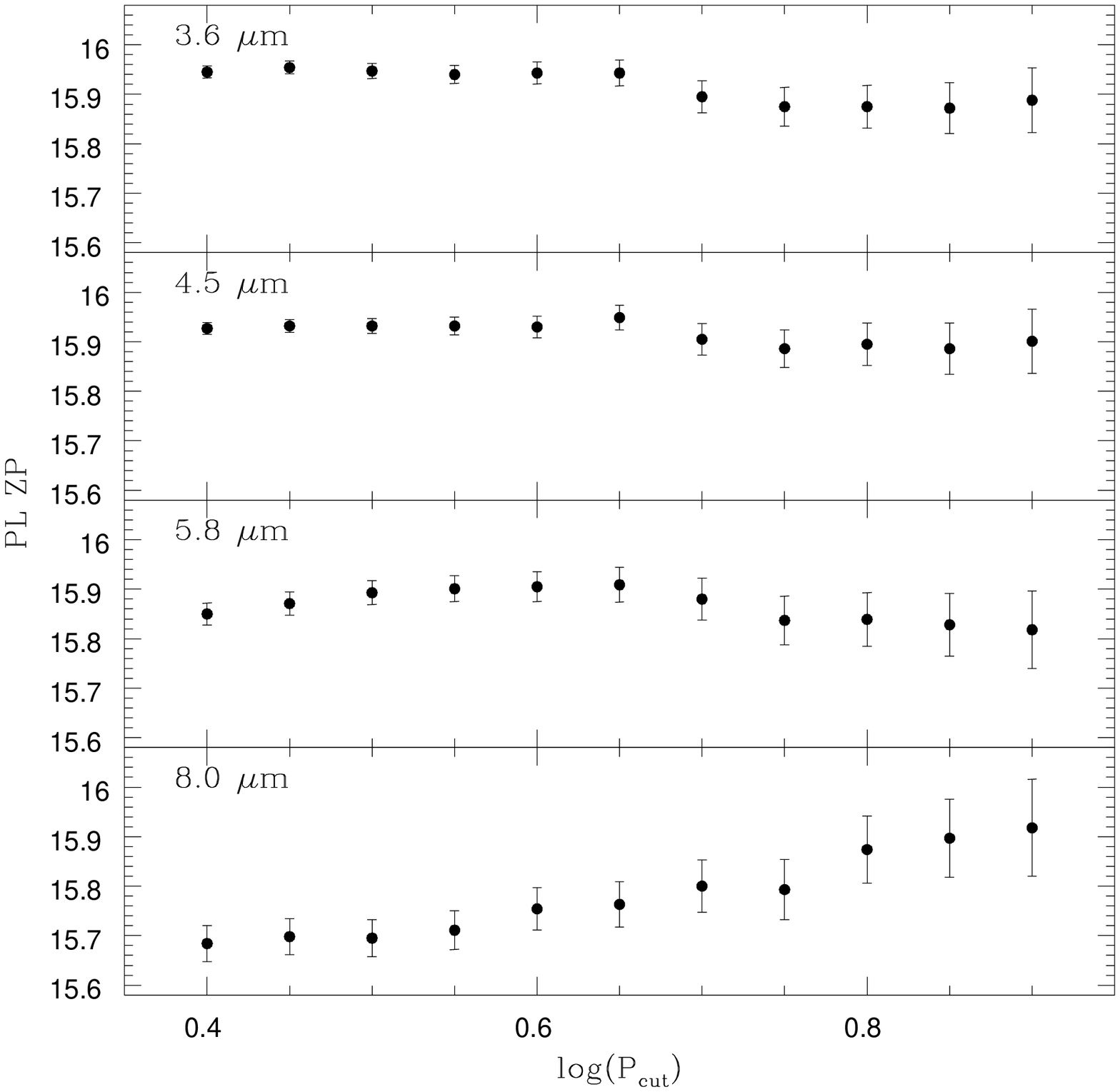}
\caption{The slopes (left panel) and the zero-points (bottom panel) of the fitted P-L relations as a function of adopted period cuts ($\log [P_{cut}]$). \label{fig_pcut}}
\end{figure*}

Even thought period cut of $\log (P)\sim0.4$ has been applied to our sample to avoid the contamination of overtone Cepheids at the short period end (see Section 2), the short period Cepheids in our sample may still be influenced by the overtone Cepheids, the increasing sensitivity of blending and the increasing of measurement errors in the {\it IRAC} band (as shown in Figure \ref{fig_mmerr}) for the faint (hence short period) Cepheids. These may bias the P-L relations when comparing to those presented in Table \ref{tab1}, and a cut at a longer period may be required to avoid the bias due to these effects. In Figure \ref{fig_pcut}, we present the slopes and zero-points of the fitted P-L relations with various period cuts from $\log (P)=0.4$ to $\log (P)=0.9$ to our sample. This figure suggests that the slopes are consistent (within the $1\sigma$ error bars) at different period cuts up to $\log (P)\sim0.65$. A similar result is also found for the zero-points. This suggests that increasing the period cuts up to $\log (P)\sim0.65$ will not greatly affect the results presented in Table \ref{tab1}, and the various sources that could bias the P-L relations may not be important. However at $\log (P)\sim0.65$, the number of Cepheids in the sample decreases to $\sim50\%$ in $3.6\mu \mathrm{m}$, $4.5\mu \mathrm{m}$ and $5.8\mu \mathrm{m}$ bands, and to $\sim23\%$ in $8.0\mu \mathrm{m}$ band. Hence we retain our results given in Table \ref{tab1} in this Paper.

\subsection{Random Phase Correction and the P-L Relations in the {\it IRAC} Band}

\begin{figure}
\plotone{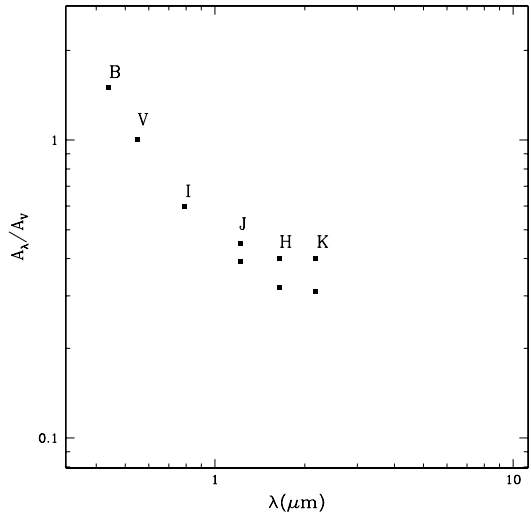}
\caption{The typical ratio of the light curve amplitude ($A$) as a function of wavelength, using the $V$ band amplitude as a reference (hence the ratio for $V$ band is 1). The amplitude ratios are from \citet{fre88} for the $B$ band, \citet{tan97} \& \citet{nge03} for the $I$ band, and \citet{sos05} for the $JHK$ band. There are two points for $JHK$ band because \citet{sos05} separated out the amplitude ratios at two period ranges. \label{fig_ratio}}
\end{figure}

In optical bands accurate mean magnitudes from the light curves are used to fit for the P-L relations. In contrast, the magnitudes obtained from the SAGE database are flux averaged over several observations \citep[see][for the details of the SAGE observing strategy]{mei06}. Since information regarding the time of observation of each data point is not available to us, these averaged magnitudes may not necessarily correspond to the mean magnitudes of the Cepheids. Nevertheless, it is well-known that the amplitude of Cepheid light curves decreases as wavelength increases. From Figure \ref{fig_ratio}, it can be seen that the amplitude decreases from $B$ to $J$ band, and flattens out at $H$ and $K$ band. Extrapolating this trend to longer wavelengths may imply that the amplitudes in the {\it IRAC} bands could either be the same as in $HK$ band, or become smaller or larger as wavelength increases. Therefore, we assume that the amplitudes will be small and these ``random-phased'' magnitudes are close to the mean magnitudes in those bands. In $JHK$ bands, there are well-developed methods to estimate the mean magnitudes from single epoch (or a few epochs) random phase observations, as presented in \citet{nik04} and \citet{sos05}. However, both methods either require the epoch of observation to be known or the existence of a template constructed from well-observed light curves. Clearly, both requirements are not present for the data we studied in this Paper. Nevertheless, the use of random-phase $JHK$ magnitudes from single epoch observations to derive the P-L relation has a precedent in the literature \citep[see, for example,][]{gro00}.

The influence of (single epoch) random phase magnitudes is expected to be minimal in the $K$ band. In order to test the validity of this assumption we use 2MASS $K$ band data that are available from the SAGE database. We repeat the fitting of the $K$ band P-L relation in the same way as for the {\it IRAC} bands. The results are presented in Table \ref{tabk}. We compare our $K$ band P-L relations with the $K$ band LMC P-L relation that is given in \citet{fou07}. There are two main reasons for selecting the P-L relation from \citet{fou07} for our comparison. Firstly, Cepheids used in \citet{fou07} are entirely based on the OGLE sample. This is similar to our sample that consists mostly of OGLE Cepheids. Secondly, random phase correction with the method described in \citet{sos05} and extinction correction have been applied to the 2MASS $K$ band data used in \citet{fou07}. This is in contrast to the data used in this Paper. 

Table \ref{tabk} reveals that without applying any random phase correction, the slope of the $K$ band P-L relation from the SAGE database is identical to the $K$ band P-L relation given in \citet{fou07}. This suggests that the random phase correction may not have a large influence on the $K$ band slope, probably due to the small amplitude in the $K$ band. The difference between the \citet{fou07} P-L relation zero points and the results found with the SAGE database is $\sim -0.06$mag. Assuming the mean extinction toward the LMC is $\sim 0.10$mag. \citep[for example, see][]{fre01} and using the total-to-selective extinction ($R$) of $0.38$ in the $K$ band \citep{fou07}, then about $-0.04$mag. of the $-0.06$mag. difference can be explained as due to the extinction. Therefore, we believe that the {\it IRAC} band P-L relations presented here will not be strongly influenced by the lack of random phase or extinction corrections.

\begin{deluxetable}{lcccc}
\tabletypesize{\scriptsize}
\tablecaption{Comparison of the $K$ Band P-L Relation.\label{tabk}}
\tablewidth{0pt}
\tablehead{
\colhead{Source} &
\colhead{Slope} &
\colhead{Zero-Point} &
\colhead{$\sigma$} &
\colhead{$N$}}
\startdata
Catalog & $-3.231\pm0.021$ & $16.050\pm0.016$ & 0.140 & 634 \\
Archive & $-3.229\pm0.021$ & $16.048\pm0.016$ & 0.141 & 642 \\
F07     & $-3.228\pm0.028$ & $15.989\pm0.006$ & 0.136 & 529 
\enddata
\tablecomments{$\sigma$ is the dispersion of the P-L relation. Catalog=SAGE Catalog; Archive=SAGE Archive; F07=\citet{fou07}.}
\end{deluxetable}

\subsection{Comparison of P-L Relations at Different Bands}

\begin{figure}
\plotone{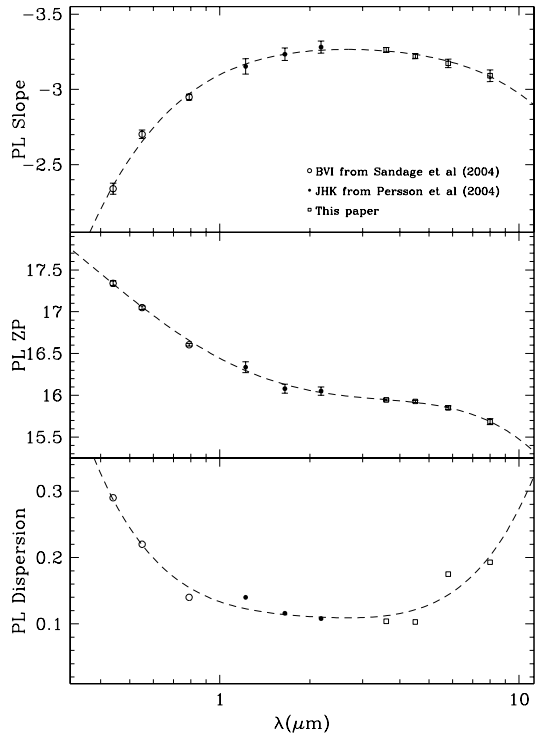}
\caption{Comparison of the slopes (top panel), the zero-points (ZP, middle panel) and the dispersions (bottom panel) for the empirical LMC P-L at different bands. The {\it IRAC} band data points are from the SAGE Archive results as given in Table \ref{tab1}. The dashed curves are the polynomial fits to the data points. \label{fig_compare}}
\end{figure}

It has been well documented in the literature that the slopes of the P-L relation become progressively steeper from $B$ to $K$ band, while at the same time the dispersion of the P-L relation decreases \citep[see, for example,][]{mad91,ber96,cap00,fio02,fio07}. This is expected in part due to the black-body curves with Cepheid-like temperatures. From $L\propto R^2T^4$, the temperature variation will dominate the luminosity variation in the optical bands \citep[see, for example,][]{cox80} and extends to $JH$ band or even the $K$ band. In contrast, at $K$ band and/or the wavelengths longer than the $K$ band, the radius variation will dominate the temperature variation. Since the period-radius (P-R) relation is independent of wavelength, the slope of the P-L relation is expected to reach a maximum value at some characteristic wavelength and remain constant as the wavelength increases. Similarly, the dispersion is expected to reach a minimum at the same characteristic wavelength and remain steady at longer wavelengths. In Figure \ref{fig_compare}, we compare the slope, the zero-point and the dispersion of the P-L relation at various bands. For illustration purposes, we adopt the empirical $BVI$ and $JHK$ band P-L relations from \citet{san04} and \citet{per04}, respectively. To extend to longer wavelengths, we also add P-L relations in the {\it IRAC} bands from Table \ref{tab1} (with results from SAGE Archive) in this figure. 

Figure \ref{fig_compare} attests to the fact that the slope of the P-L relation is the steepest around the $K$ to $4.5\mu \mathrm{m}$ band region, and becomes shallower at longer wavelengths. In contrast, the zero-point of the P-L relation is a monotonic function of wavelength. The dispersion of the P-L relation displays a similar trend to the slope: the dispersion reaches a minimum around the $K$ and $4.5\mu \mathrm{m}$ bands, and subsequently increases for longer wavelengths. A polynomial function in the form of $Y(\lambda) = a_0 + \sum^{i=4}_{i=1} a_i \times [\log_{10}(\lambda)]^i$ was used to fit the data points in Figure \ref{fig_compare}. Here $Y$ represents either the P-L slopes, zero-points or dispersion. From the polynomial fits, the steepest P-L slope of $-3.266$ seems to be located at a wavelength of $\sim 2.7$ microns. The minimum dispersion of $\sim0.109$ occurs at the same wavelength. This suggests that the P-L relations from $K$ band to $3.6\mu \mathrm{m}$ band will provide a more accurate distance scale measurement than the optical bands. However, observations from optical bands are still required to detect Cepheid variables. Furthermore, the bottom panel of Figure \ref{fig_compare} implies that the dispersions at $3.6\mu \mathrm{m}$ and $4.5\mu \mathrm{m}$ bands are similar to the $JHK$ bands, suggesting also that omission of random phase corrections may not be too serious an oversight.

\begin{figure}
\plotone{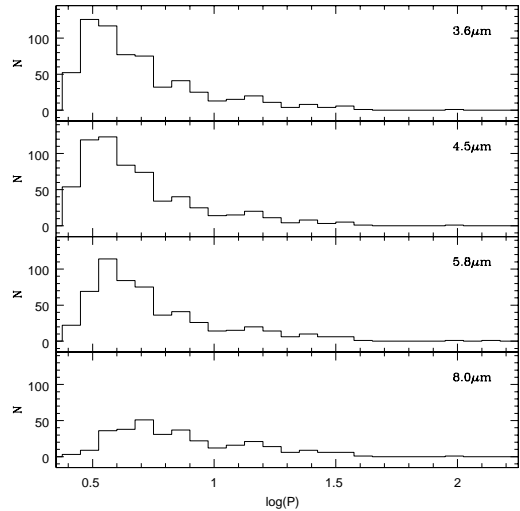}
\caption{The period distribution for the SAGE Archive samples that used to derive the P-L relation given in Table \ref{tab1}. \label{fig_period}}
\end{figure}

Figure \ref{fig_compare} provides evidence that the slopes and dispersions for the $3.6\mu \mathrm{m}$ and $4.5\mu \mathrm{m}$ band P-L relations are consistent both with each other and with the $K$ band P-L relation. However, the slopes and dispersions for the $5.8\mu \mathrm{m}$ and $8.0\mu \mathrm{m}$ band P-L relations are shallower and larger, respectively than the theoretical expectation outlined previously. There are two possible causes for this. The first possibility is that measurement errors become larger toward the faint end of the $5.8\mu \mathrm{m}$ and $8.0\mu \mathrm{m}$ band P-L relations, as suggested in Figure \ref{fig_mmerr}. The second possibility is that the number of Cepheids in these two bands is less than those in the $3.6\mu \mathrm{m}$ and $4.5\mu \mathrm{m}$ bands. This is especially true for the $8.0\mu \mathrm{m}$ band. Figure \ref{fig_period} presents the period distribution for our Cepheid sample: fewer Cepheids are indeed detected for $\log (P)\aplt 0.8$ in the $8.0\mu \mathrm{m}$ band. This is also reflected in Table \ref{tab1} where the number of Cepheids in the $8.0\mu \mathrm{m}$ band (from the SAGE Archive), is about 43\% to 50\% less than the number of Cepheids in other bands. \citet{mei06} reported the limiting magnitude in each band as $m^{\mathrm{limit}}_{3.6,\ 4.5,\ 5.8,\ 8.0}=\{18.348,\ 17.474,\ 15.164,\ 14.226\}$mag. This may result in fewer detections around $m\sim14$mag. in the $8.0\mu \mathrm{m}$ band. In the case of the $5.8\mu \mathrm{m}$ band, Figure \ref{fig_period} also suggests that, for $\log (P) < 0.5$, fewer Cepheids are detected when compared to the $3.6\mu \mathrm{m}$ and $4.5\mu \mathrm{m}$ bands.

\subsection{Nonlinearity of the P-L Relations \label{sec33}}

\begin{deluxetable*}{lcccccccccc}
\tabletypesize{\scriptsize}
\tablecaption{$F$-Test Results of the P-L Relations in {\it IRAC} Band.\label{tab2}}
\tablewidth{0pt}
\tablehead{
\colhead{} & \multicolumn{4}{c}{$P<10$ days} & \multicolumn{4}{c}{$P>10$ days} & \colhead{} & \colhead{} \\
\colhead{Band} &
\colhead{Slope$_S$} &
\colhead{Zero-Point$_S$} &
\colhead{$\sigma_S$} &
\colhead{$N_S$} &
\colhead{Slope$_L$} &
\colhead{Zero-Point$_L$} &
\colhead{$\sigma_L$} &
\colhead{$N_L$} &
\colhead{$F$} &
\colhead{$p(F)$} 
}
\startdata
$3.6\mu \mathrm{m}$  & $-3.309\pm0.030$ & $15.971\pm0.019$ & 0.101 & 550 & $-3.287\pm0.079$ & $15.986\pm0.099$ & 0.126 & 78 & 1.78 & 0.170 \\
$4.5\mu \mathrm{m}$  & $-3.255\pm0.029$ & $15.946\pm0.019$ & 0.100 & 558 & $-3.331\pm0.079$ & $16.076\pm0.098$ & 0.122 & 77 & 2.81 & 0.061 \\
$5.8\mu \mathrm{m}$  & $-3.121\pm0.058$ & $15.817\pm0.039$ & 0.176 & 471 & $-3.212\pm0.093$ & $15.893\pm0.118$ & 0.175 & 90 & 0.55 & 0.579 \\
$8.0\mu \mathrm{m}$  & $-2.858\pm0.094$ & $15.517\pm0.069$ & 0.191 & 232 & $-3.364\pm0.114$ & $16.018\pm0.143$ & 0.186 & 87 & 5.90 & 0.003    
\enddata
\tablecomments{$\sigma$ is the dispersion of the P-L relation.}
\end{deluxetable*}

A number of recent studies have strongly suggested the LMC P-L relation in {\it optical} bands is nonlinear, in the sense that the relation can be broken into two P-L relations separated at/around 10 days \citep{tam02,kan04,san04,kan06,nge05,nge06a,nge06b,kan07b,nge07b}. In NIR, \citet{nge05} and \citet{nge07b} found that the $JH$ band LMC P-L relations are nonlinear but the $K$ band LMC P-L relation is marginally linear. \citet{nge06b} outline a black-body argument for their result that the P-L relation could be linear in $K$ band but not in the optical and/or $JH$ band. One possibility is that the temperature variation in Cepheid atmospheres, modulated at certain phases, periods and metallicities by the stellar photosphere-hydrogen ionization front interaction \citep{kan04b,kan06,kan07a}, is responsible for the observed nonlinear P-L relation. Since the temperature variation for a black-body with Cepheid-like temperatures is minimal or even negligible at longer wavelengths, it is expected that the P-L relation becomes linear for wavelengths longer than the $K$ band. Therefore, another main motivation for this Paper is to study the linearity/nonlinearity of the LMC P-L relations in the {\it IRAC} band.

To test the nonlinearity of the P-L relations, we apply the $F$-test as in our previous studies \citep{kan04,kan06,nge05}. The detailed description and formalism of the $F$-test can be found in \citet{wei80}, \citet{kan04} and \citet{nge05}, and will not be repeated here. Simply speaking, in our $F$-test, the null hypothesis is that the data can be fitted with a single regression line, and the alternate hypothesis is that two regression lines separated at 10 days are needed to fit the data. In our test, we set $p(F)$, the probability of the observed $F$ value under the null hypothesis, to be $0.05$ (equivalently at the 95\% confidence level). This corresponds to $F\sim3$ for our data. Hence, $F>3$ indicates that the null hyopthesis can be rejected at the 95\% confidence level or more for the P-L relation under scrutiny. 

In Table \ref{tab2}, we present the results from the $F$-test for the P-L Relations in {\it IRAC} band. The $F$-test results indicate that the P-L relation is linear in the $3.6\mu \mathrm{m}$, $4.5\mu \mathrm{m}$ and $5.8\mu \mathrm{m}$ band but not in the $8.0\mu \mathrm{m}$ band. The linearity of the P-L relation in $3.6\mu \mathrm{m}$, $4.5\mu \mathrm{m}$ and $5.8\mu \mathrm{m}$ band is expected from the black-body argument as outlined previously. However it is important to point out that the linearity/nonlinearity of the P-L relation in these bands does not necessary imply the P-L relation in the optical bands will be linear/nonlinear. The apparently nonlinear $8.0\mu \mathrm{m}$ P-L relation is puzzling. Removing the longest period Cepheid in the $8.0\mu \mathrm{m}$ band sample still leaves a nonlinear result with a $F$ value of $5.04$. The relatively small number of Cepheids in $8.0\mu \mathrm{m}$ band (see Figure \ref{fig_period}) may cause the apparent nonlinear result though the $F$ test is sensitive to this. The similar trends of the $1\sigma$ measurement error plot, as shown in Figure \ref{fig_mmerr}, and the similar dispersions for the P-L relations in the $5.8\mu \mathrm{m}$ and $8.0\mu \mathrm{m}$ band, suggest the lack of short period Cepheids in $8.0\mu \mathrm{m}$ band may be the reason for the nonlinear result, because the $F$-test result finds that the $5.8\mu \mathrm{m}$ P-L relation is linear. It is still inconclusive if the $8.0\mu \mathrm{m}$ P-L relation is truely nonlinear or not and more data are needed in the future work to solve this problem.

\section{The Period-Color Relation, the Color-Color Plot and the CMD}

\begin{figure*}
\plottwo{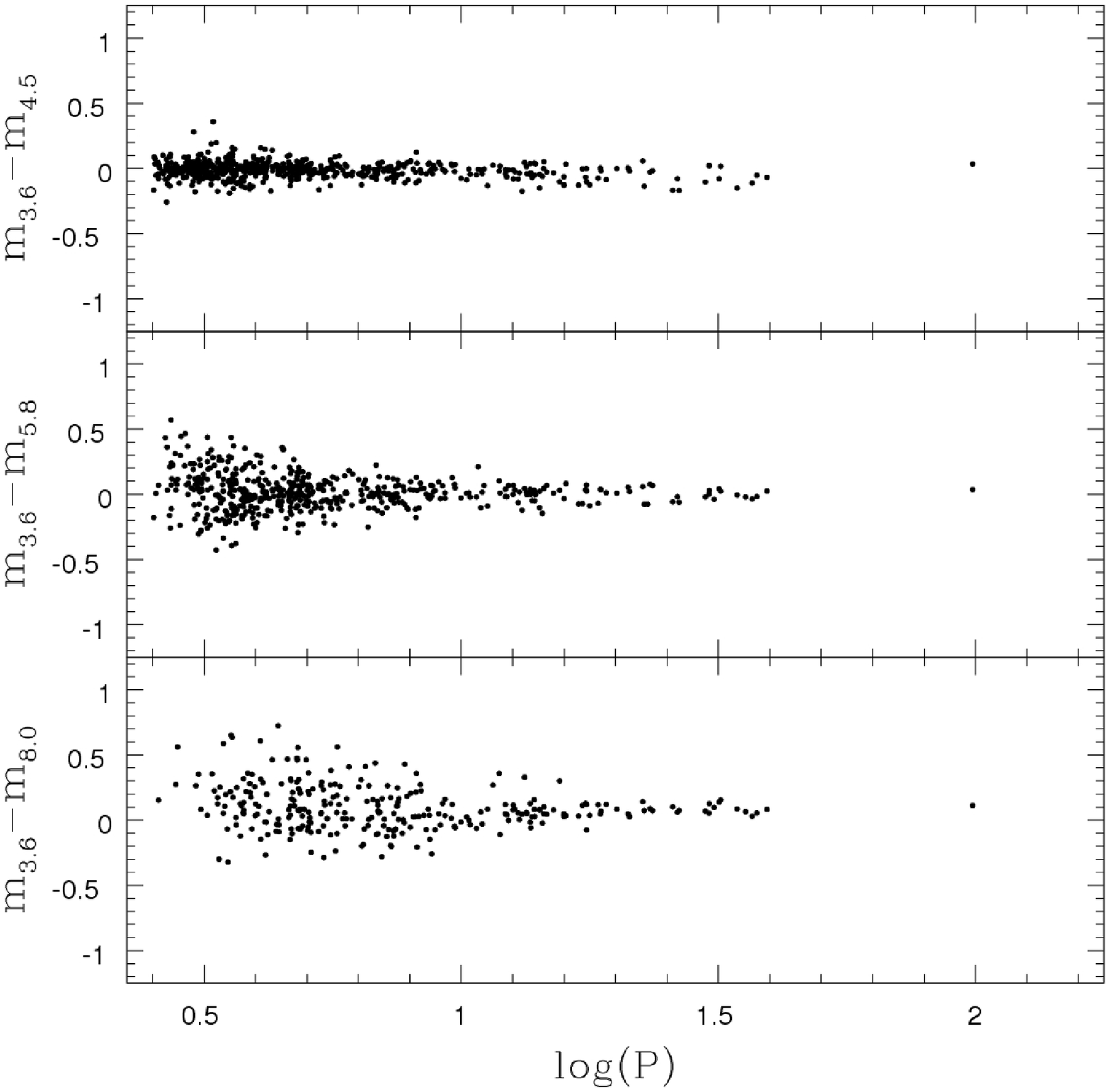}{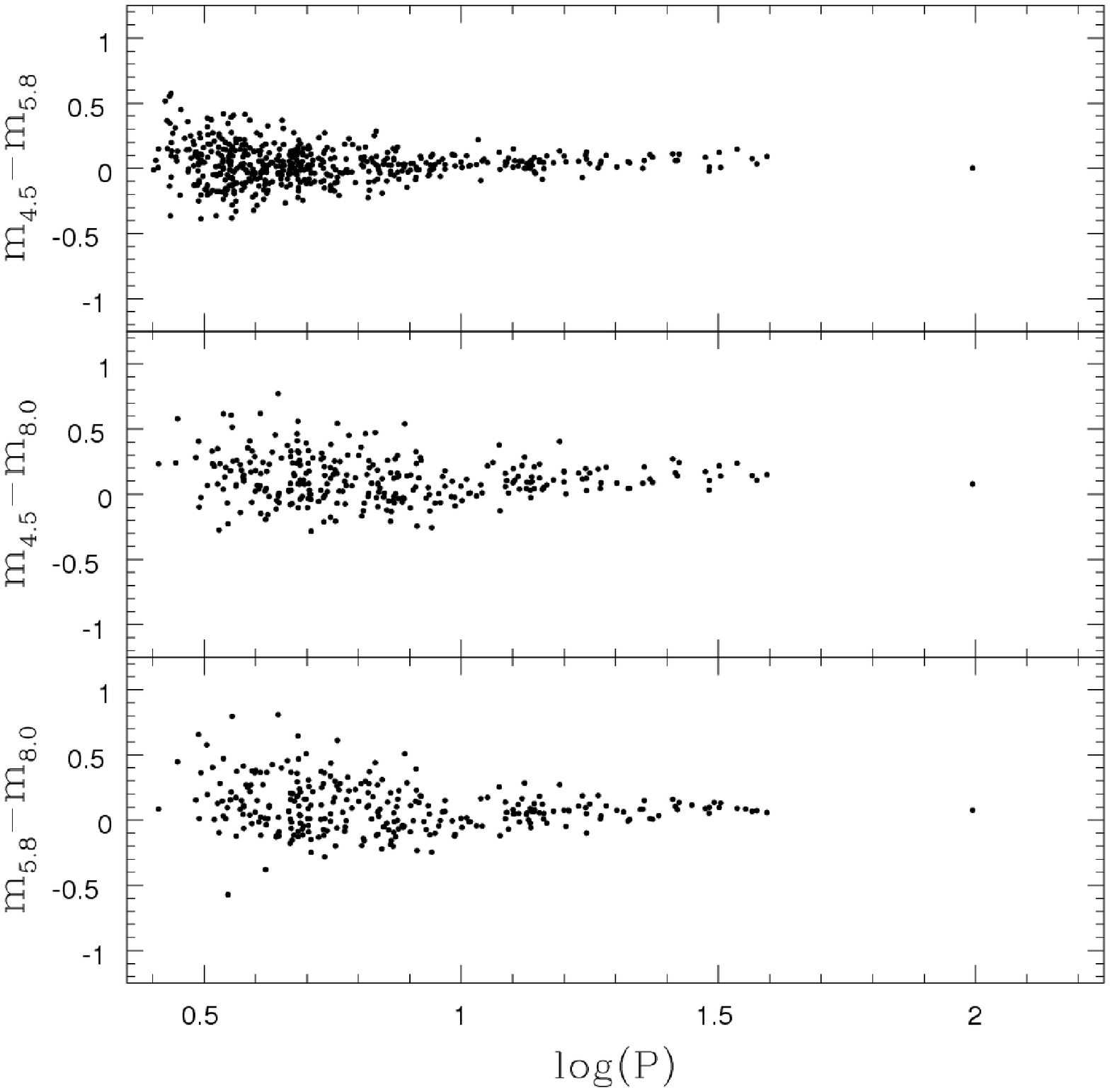}
\caption{The P-C relations from the matched sources in the SAGE Archive after the removal of the outliers. \label{fig_pc}}
\end{figure*}

\begin{figure}
\plotone{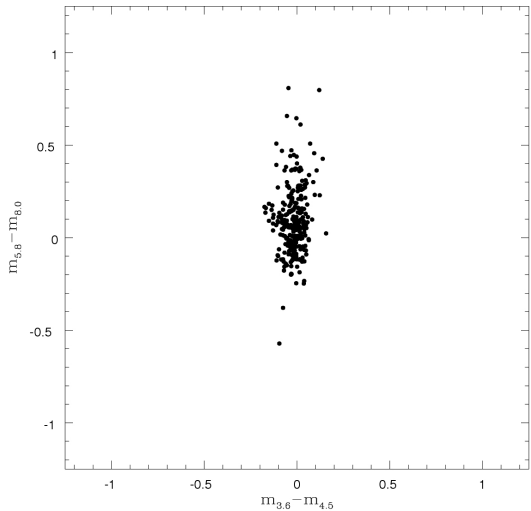}
\caption{The color-color plot from the 261 matched sources in the SAGE Archive, after the removal of the outliers, with detections in all four bands. \label{fig_cc}}
\end{figure}

\begin{figure*}
\plottwo{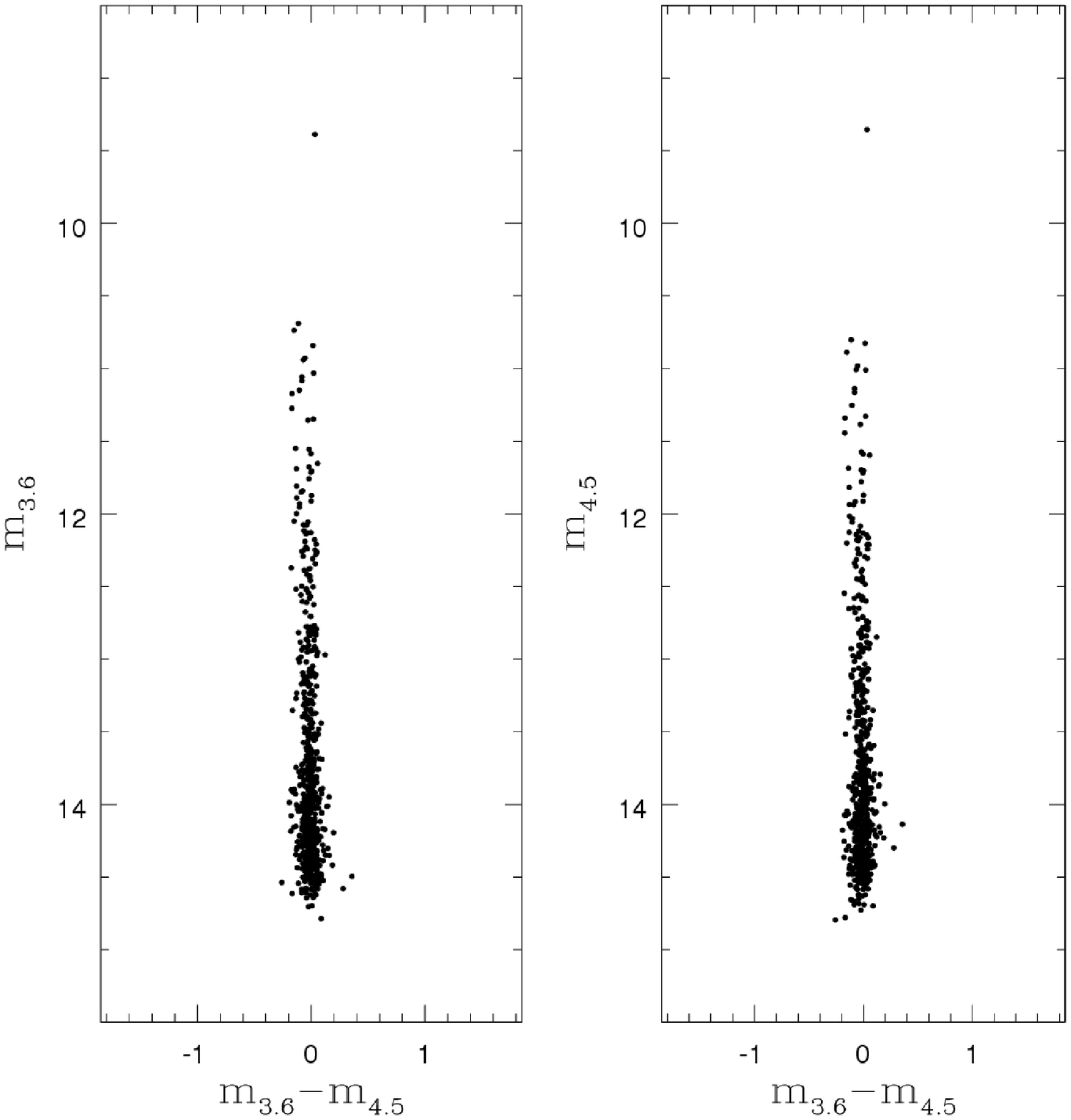}{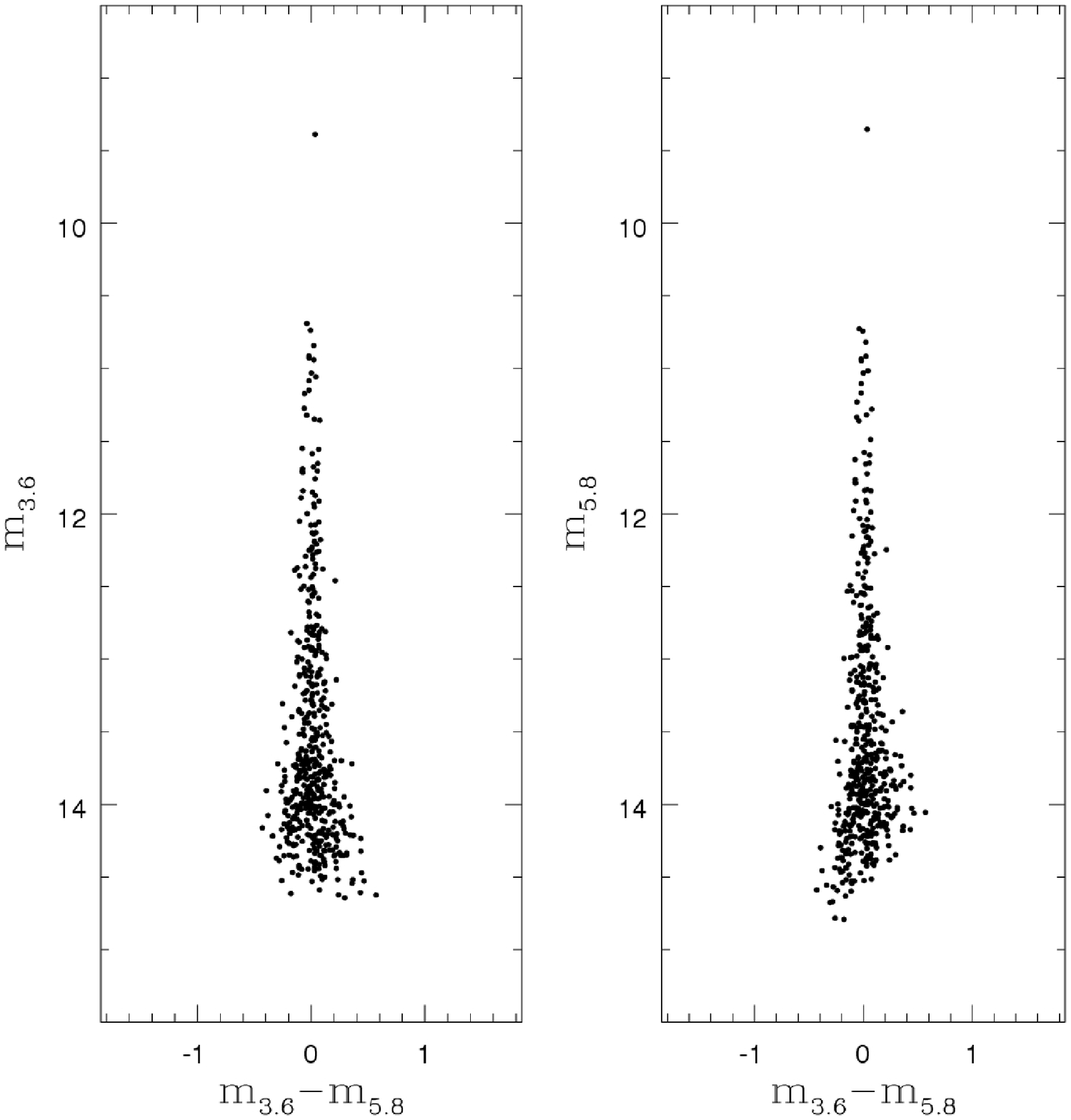} \\
\plottwo{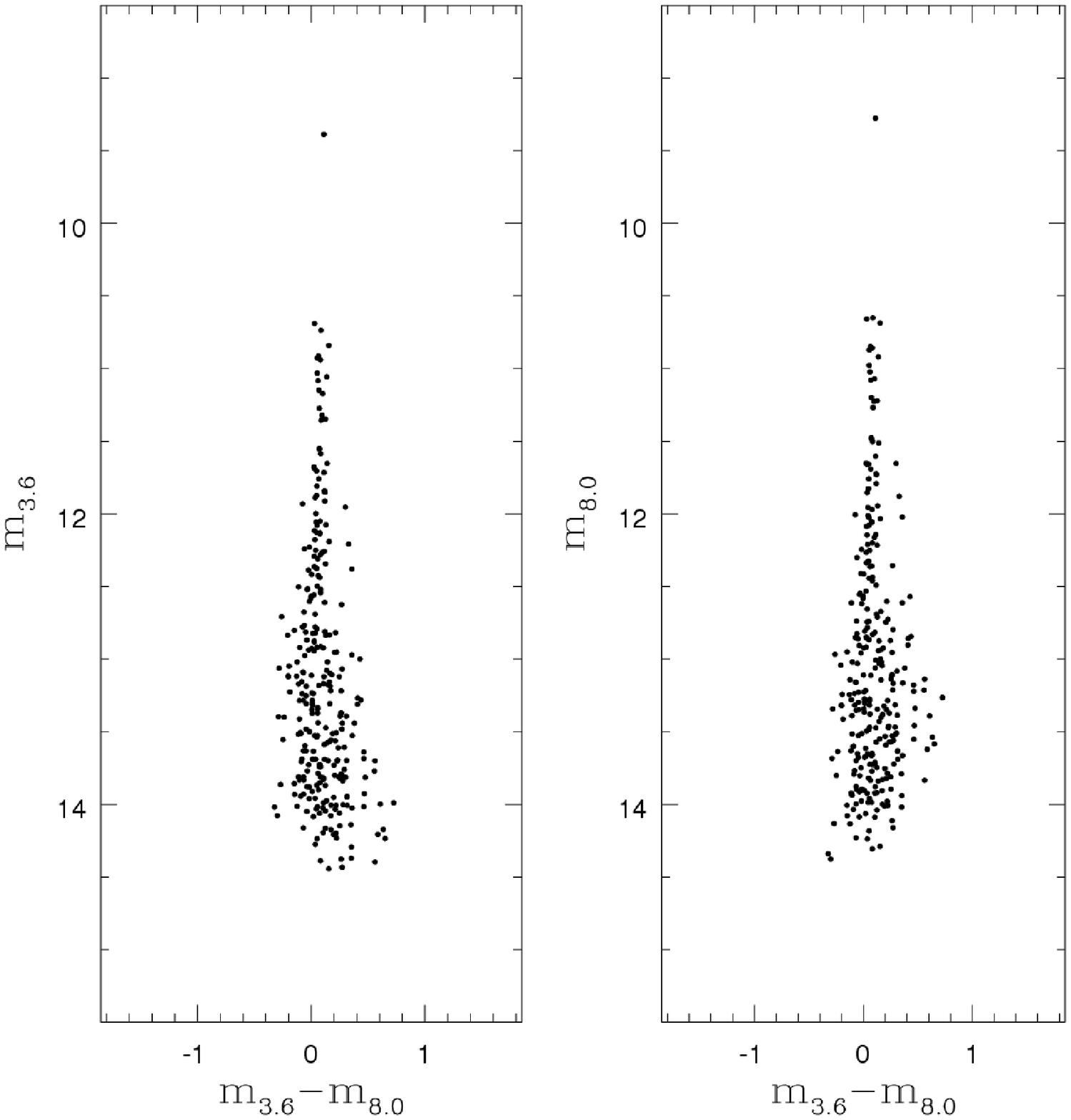}{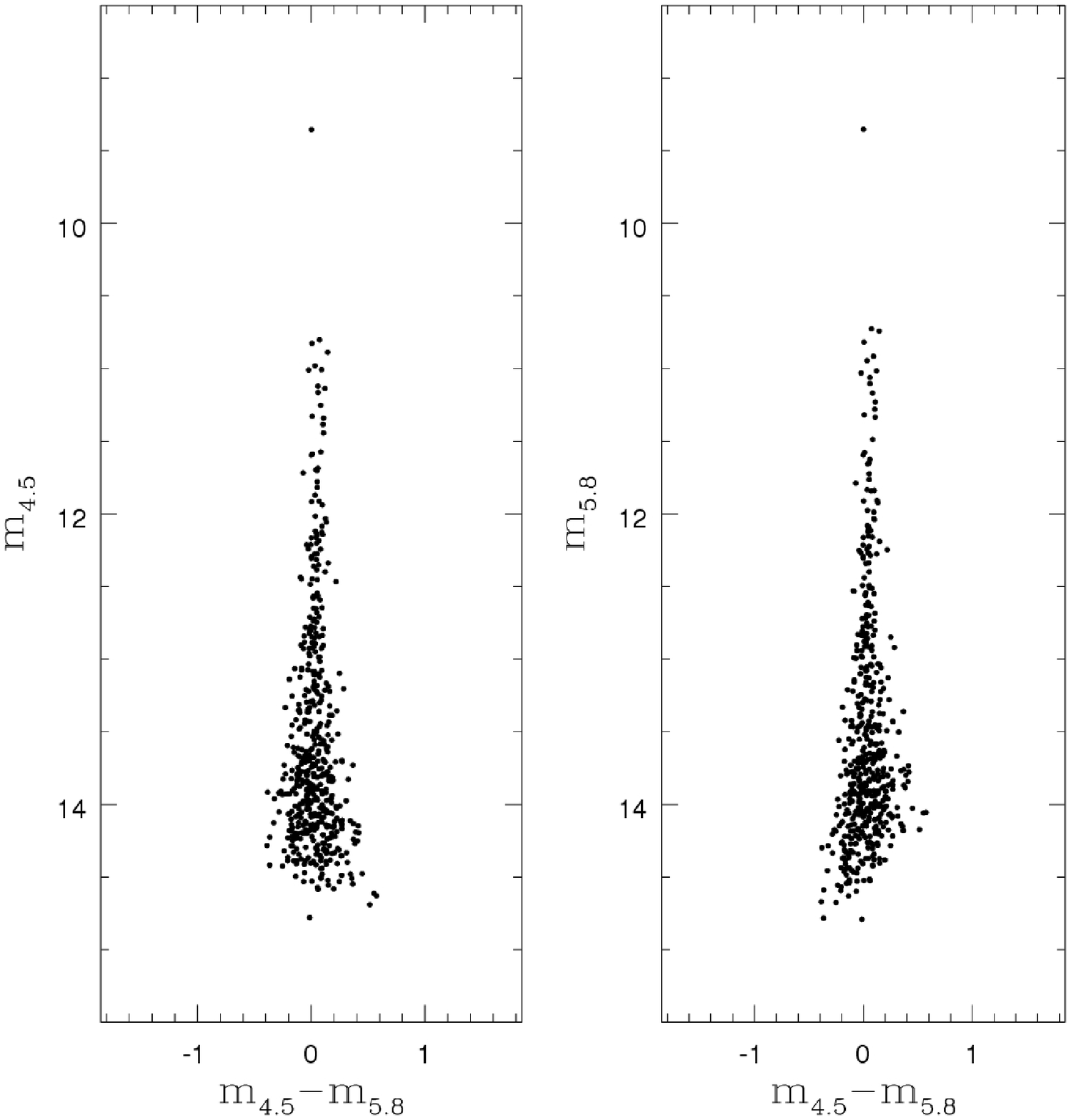} \\
\plottwo{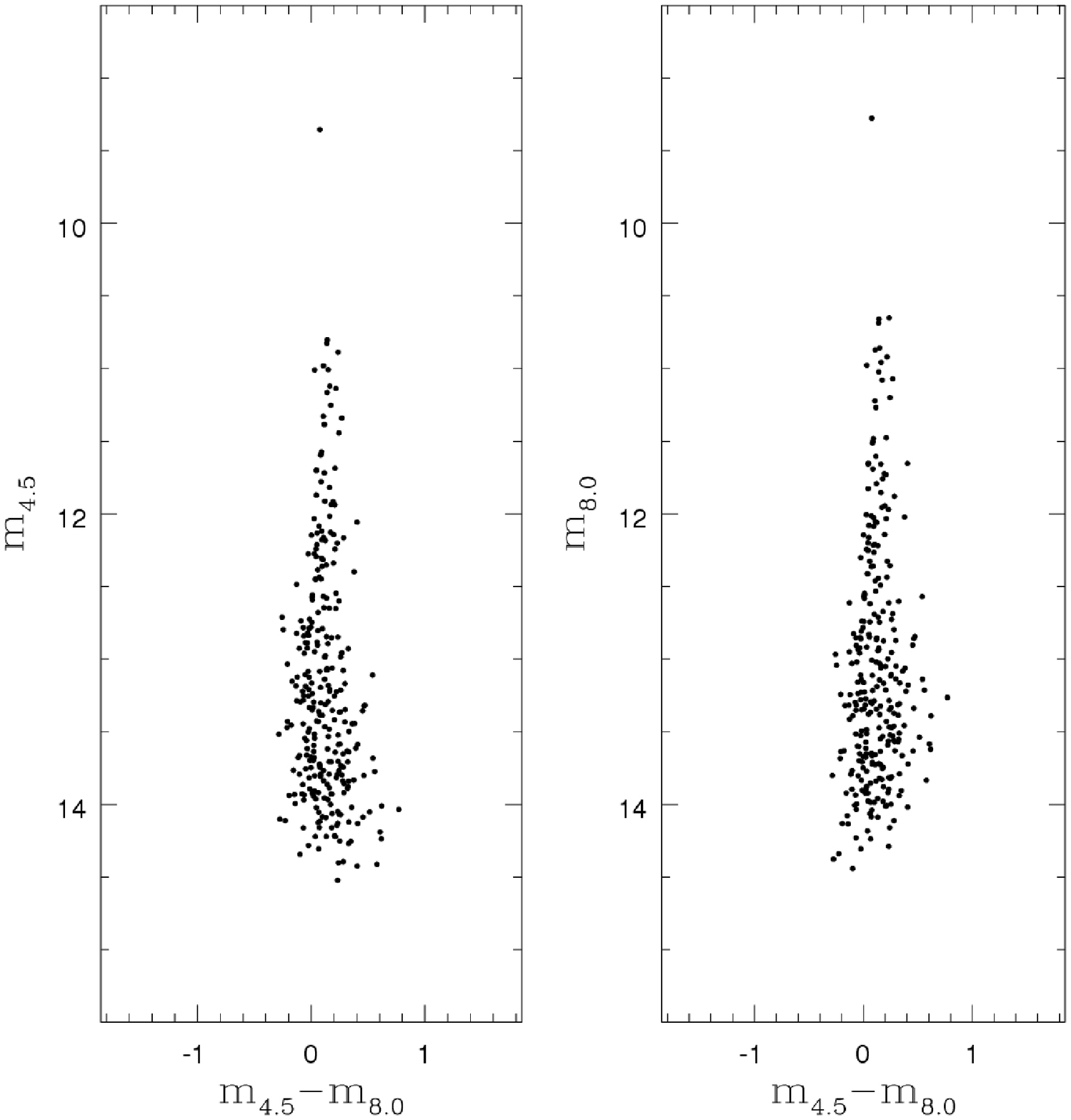}{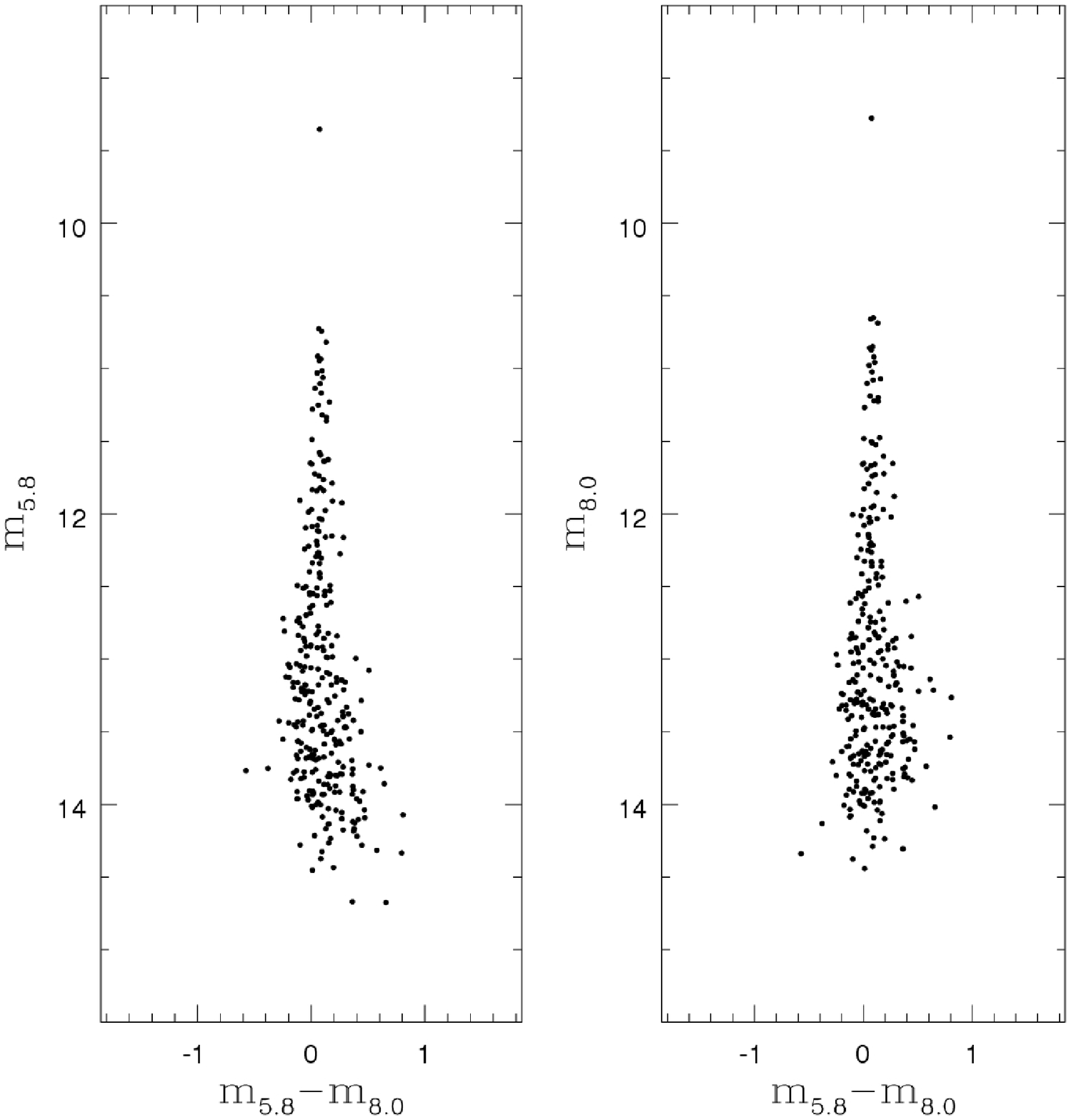} 
\caption{The CMD from the matched sources in the SAGE Archive after the removal of the outliers.  \label{fig_cmd}}
\end{figure*}

In addition to the P-L relations derived in previous sections, the SAGE Archive also permits the derivation of P-C relations, color-color plots and the CMD. The resulting P-C relations from the data are presented in Figure \ref{fig_pc} and Table \ref{tab3}, while the color-color plots and CMD are presented in Figures \ref{fig_cc} and \ref{fig_cmd}, respectively. 

The {\it IRAC} band P-C relations are found to be relatively flat as compared to the P-C relations in the optical band, with the mean color being close to zero (especially for the $m_{4.5}-m_{5.8}$ P-C relation). This is not a surprise given that the slopes and the zero-points of the P-L relations in these bands are similar (see Table \ref{tab1} and Figure \ref{fig_compare}). In fact black-body curves with Cepheid-like temperatures predict the P-C relation should vanish at these bands. Table \ref{tab3} also finds that some P-C relations are identical to each other, including the $m_{3.6}-m_{8.0}$ \& $m_{5.8}-m_{8.0}$ pair, and the $m_{3.6}-m_{4.5}$, $m_{3.6}-m_{5.8}$ \& $m_{5.8}-m_{8.0}$ P-C relations. Furthermore, all P-C relations presented in Figure \ref{fig_pc} display a very tight sequence for Cepheids with $\log(P) \apgt 1.0$. This is also seen in the CMD. In Table \ref{tabpc} we present the P-C relations separated at 10 days. The flatness of the P-C slope, the mean color of zero and the small dispersion of the P-C relation are clearly evident from this table for Cepheids with period longer than 10 days.

\begin{deluxetable}{lcccc}
\tabletypesize{\scriptsize}
\tablecaption{P-C Relations in {\it IRAC} Band.\label{tab3}}
\tablewidth{0pt}
\tablehead{
\colhead{Color} &
\colhead{Slope} &
\colhead{Zero-Point} &
\colhead{$\sigma$} &
\colhead{$N$} }
\startdata
$m_{3.6}-m_{4.5}$  & $-0.044\pm0.010$ & $0.017\pm0.007$ & 0.061 & 595 \\
$m_{3.6}-m_{5.8}$  & $-0.048\pm0.024$ & $0.050\pm0.019$ & 0.137 & 504 \\
$m_{3.6}-m_{8.0}$  & $-0.108\pm0.038$ & $0.190\pm0.034$ & 0.171 & 287 \\ 
$m_{4.5}-m_{5.8}$  & $-0.008\pm0.025$ & $0.039\pm0.019$ & 0.140 & 512 \\
$m_{4.5}-m_{8.0}$  & $-0.057\pm0.038$ & $0.169\pm0.034$ & 0.170 & 291 \\
$m_{5.8}-m_{8.0}$  & $-0.104\pm0.038$ & $0.186\pm0.035$ & 0.181 & 298 
\enddata
\tablecomments{$\sigma$ is the dispersion of the P-C relation.}
\end{deluxetable}

\begin{deluxetable*}{lcccccccc}
\tabletypesize{\scriptsize}
\tablecaption{The P-C Relations in {\it IRAC} Band Separated at 10 days.\label{tabpc}}
\tablewidth{0pt}
\tablehead{
\colhead{} & \multicolumn{4}{c}{$P<10$ days} & \multicolumn{4}{c}{$P>10$ days} \\
\colhead{Band} &
\colhead{Slope$_S$} &
\colhead{Zero-Point$_S$} &
\colhead{$\sigma_S$} &
\colhead{$N_S$} &
\colhead{Slope$_L$} &
\colhead{Zero-Point$_L$} &
\colhead{$\sigma_L$} &
\colhead{$N_L$} } 
\startdata
$m_{3.6}-m_{4.5}$  & $-0.017\pm0.019$ & $0.001\pm0.012$ & 0.061 & 521 & $-0.031\pm0.039$ & $-0.007\pm0.048$ & 0.059 & 74 \\ 
$m_{3.6}-m_{5.8}$  & $-0.117\pm0.050$ & $0.093\pm0.034$ & 0.146 & 429 & $-0.012\pm0.037$ & $ 0.018\pm0.046$ & 0.058 & 75 \\ 
$m_{3.6}-m_{8.0}$  & $-0.370\pm0.097$ & $0.376\pm0.072$ & 0.189 & 213 & $ 0.057\pm0.051$ & $ 0.001\pm0.064$ & 0.079 & 74 \\ 
$m_{4.5}-m_{5.8}$  & $-0.118\pm0.051$ & $0.107\pm0.034$ & 0.149 & 438 & $ 0.027\pm0.035$ & $ 0.015\pm0.043$ & 0.053 & 74 \\ 
$m_{4.5}-m_{8.0}$  & $-0.335\pm0.094$ & $0.364\pm0.070$ & 0.185 & 217 & $ 0.096\pm0.060$ & $-0.000\pm0.075$ & 0.092 & 74 \\ 
$m_{5.8}-m_{8.0}$  & $-0.377\pm0.106$ & $0.381\pm0.078$ & 0.204 & 215 & $ 0.075\pm0.047$ & $-0.024\pm0.059$ & 0.076 & 83    
\enddata
\tablecomments{$\sigma$ is the dispersion of the P-C relation.}
\end{deluxetable*}

At the short period end of the P-C relations, the dispersion of the P-C relations gets broader (except for the $m_{3.6}-m_{4.5}$ P-C relation) as period decreases. This feature is also seen from Figure \ref{fig_cmd} toward the faint end of the CMD. Probably this could be due to the relatively large measurement errors at the faint (or short period) end, because the measurement errors in color can reach up to $\sim0.3$mag. as suggested from Figure \ref{fig_mmerr}. In addition, Figures \ref{fig_pc} and \ref{fig_cmd} imply a lack of detections for the matched sources near $\log(P)\sim0.5$, especially for the P-C relations that include the $8.0\mu \mathrm{m}$ band. This could cause the short period P-C relations to deviate from flatness and mean zero color as given in Table \ref{tabpc}. Because of these reasons, we did not test the nonlinearity of the P-C relations with the $F$-test. 

Nevertheless, the CMD presented in Figure \ref{fig_cmd} finds that the instability strip for LMC Cepheids is well defined in the {\it IRAC} band, especially with the $m_{3.6}-m_{4.5}$ color. The tightness of the $m_{3.6}-m_{4.5}$ color is also reflected in Figure \ref{fig_cc}, where the spread out of $m_{5.8}-m_{8.0}$ color is mainly from the short period Cepheids. The well occupied regions of the Cepheids in Figure \ref{fig_cc} and the well defined CMD suggest that Figure \ref{fig_cc} and \ref{fig_cmd} can be used to identify Cepheids in future studies.

\section{Conclusion}

In this Paper, we derive P-L relations for LMC Cepheids in {\it IRAC} $3.6$, $4.5$, $5.8$ and $8.0$ microns bands. These P-L relations can be potentially applied to future extragalactic distance scale studies with, for example, the {\it JWST}. The data are taken from the {\it Spitzer's} archival database from the SAGE program. After properly removing the outliers, the fitted P-L relations are presented in Table \ref{tab1}. We have tested the P-L relations with various period cuts and found that our results are insensitive to period cuts up to $\log P_{cut} \sim 0.65$. We also argue that the random phase corrections may not be important for {\it IRAC} band P-L relations. When comparing P-L relations from $B$ to $8.0\mu \mathrm{m}$ bands, the slope of the P-L relation appears to be the steepest around $K$ band to $3.6\mu \mathrm{m}$ band, while the dispersion of the P-L relation reaches a minimum between those bands. The shallower slopes and larger P-L dispersions in the $5.8\mu \mathrm{m}$ and $8.0\mu \mathrm{m}$ band are in contrast to the theoretical expectation. This could be due to the smaller number of Cepheids and larger measurement errors toward the faint end in these two bands. We also test the nonlinearity of the P-L relations in the {\it IRAC} band using the $F$ statistical test. As expected, the $F$-test results show that the P-L relations are linear in $3.6$, $4.5$ and $5.8$ microns band, but the $8.0\mu \mathrm{m}$ P-L relation is found to be nonlinear. However, the nature of the nonlinear $8.0\mu \mathrm{m}$ P-L relation is still inconclusive. For the P-C relations, it was found that the slopes of the P-C relation are relatively flat in the {\it IRAC} bands. Finally, the LMC Cepheids show a well-defined instability strip in the CMD and clustered in a small region in the color-color plot. This can potentially be used to identify Cepheids observed in the {\it IRAC} bands. Even though there may be some associated problems for the $8.0\mu \mathrm{m}$ P-L relation, the $3.6\mu \mathrm{m}$, $4.5\mu \mathrm{m}$ and perhaps the $5.8\mu \mathrm{m}$ P-L relations can still be used in future distance scale studies.

\acknowledgements
We thank the anonymous referee for helpful suggestions to improve the manuscript, and Margaret Meixner for clearing out some issues regarding the SAGE database. We also thank Lucas Macri and Nancy Evans for useful discussion. CN acknowledges support from NSF award OPP-0130612 and a University of Illinois seed funding award to the Dark Energy Survey. SMK acknowledges support from the Chretien International Research Award from the American Astronomical Society. This research has made use of the NASA/IPAC Infrared Science Archive, which is operated by the Jet Propulsion Laboratory, California Institute of Technology, under contract with the National Aeronautics and Space Administration.

\end{document}